\documentclass[a4paper, 11pt]{article}
\usepackage[margin=1.1in]{geometry}
\usepackage{amsmath,amsfonts,amsthm}
\usepackage{graphicx,cite,times}
\usepackage{enumitem,version}
\usepackage{pdfpages,hyperref}

\newtheorem{theorem}{Theorem}[section]

\newtheorem{lemma}[theorem]{Lemma}

\newtheorem{example}{Example}

\newtheorem{definition}{Definition}[section]

\numberwithin{equation}{section}
\Large

\usepackage{tikz,xcolor,hyperref}
\definecolor{lime}{HTML}{A6CE39}
\DeclareRobustCommand{\orcidicon}{\hspace{-1mm}
	\begin{tikzpicture}
		\draw[lime, fill=lime](0,0) circle[radius=0.16] 
		node[white] {{\fontfamily{qag}\selectfont \tiny ID}};
		\draw[white, fill=white] (-0.0625,0.095) 
		circle [radius=0.007];
	\end{tikzpicture}
	\hspace{-1mm}}
\foreach \x in {A, ..., Z}{\expandafter\xdef\csname orcid\x\endcsname{\noexpand\href{https://orcid.org/\csname orcidauthor\x\endcsname}
		{\noexpand\orcidicon}}}
\begin{document}
	\title{A public key cryptography using multinacci block matrices}
	\author{Munesh Kumari$^{1}$\footnote{E-mail: muneshnasir94@gmail.com}\orcidA{} and Jagmohan Tanti\footnote{E-mail: jagmohan.t@gmail.com}\orcidC{}
		\\\normalsize{$^{1}$\small Department of Mathematics, Central University of Jharkhand, India, 835205} 
		\\\normalsize{$^{2}$\small Department of Mathematics, Babasaheb Bhimrao Ambedkar University, Lucknow, India, 226025}    }    
	\date{\today}
	\maketitle
	\noindent\rule{15cm}{.15pt}
	\begin{abstract}
		 In this paper, we have proposed a public key cryptography using recursive block matrices involving generalized Fibonacci numbers over a ﬁnite ﬁeld $\mathbb{F}_{p}$. For this, we define multinacci block matrices, a type of upper triangular matrix involving multinacci matrices at diagonal places and obtained some of its algebraic properties. Moreover, we have set up a method for key element agreement at end users, which makes the cryptography more efficient. The proposed cryptography comes with a large key space and its security relies on the Discrete Logarithm Problem(DLP).
	\end{abstract}
	\noindent\rule{15cm}{.15pt}
	\section{Introduction}
	Information is one of the most valuable assets since the dawn of civilizations. The secured transmission of information is of prime importance. Cryptography is the science of study about the security, privacy, and conﬁdentiality of information transmitted over a secured channel. 
	The problem concerned to the topic is to study a public key cryptography with the reduction in complexity for key generation without compromising the security.
	As an example, for the same level of security, the RSA cryptosystem uses a bigger key size than the key size of Elliptic curve cryptography.
	
	In 1976, Difﬁe and Hellman\cite{diffie1976new}  provided a solution to the long-standing problem of key exchange and pointed the way to a digital signature. In 1978 Rivest, Shamir \& Adleman \cite{rivest1978method} proposed a public key cryptosystem which famed as RSA cryptosystem. The security of RSA cryptosystem depends on the difﬁculty level of factoring large integers.
	
	R.ALVAREZA, et.al \cite{alvarez2007new} proposed a public key cryptosystem (PKC) based on the generalization of the discrete logarithm problem for block matrices over the ﬁeld $\mathbb{Z}_{p}$ with the reduced key length for a given level of security. Kuppuswamy, et.al \cite{kuppuswamy2014hybrid} has given two different types of encryption algorithms, one of them is public key cryptography based on a linear block cipher and other one is private key cryptography based on a simple symmetric algorithm. Viswanath and Kumar \cite{viswanath2015public} proposed a public key cryptosystem using Hill’s cipher, in which security of the system depends on the involvement of two digital signatures. To reduce complexity and enhance processing of key in cryptography K. Prasad and H. Mahato \cite{prasad2021cryptography} have proposed a public key cryptography using generalized Fibonacci matrices in which one have to send numbers instead of matrices for key. M. Zeriouh, et.al \cite{zeriouh2019cryptography}, proposed the concept of key exchange between Alice and Bob using specially designed matrices. In this key exchange scheme, each of the sender and receiver ﬁrst chooses a square matrix of suitable order and then both publish their corresponding set of matrices that commute with their corresponding chosen matrices. Some recent work on study of cryptography using number sequences, associated matrices and classical cryptography can be seen in\cite{prasad2021cryptography,prasad2022novel,viswanath2015public,sundarayya2019public,mohan2016improved}.
	
	This paper is organized as follows. In section 2, we have first defined the multinacci block matrices and established its some properties then we have proposed a public key cryptography using extended Hill's cipher and give a key agreement method for end users. In section 3, we have illustrated the scheme by a numerical example. And lastly in section 4, we have analyzed the keyspace and mathematical strength of the scheme followed by conclusion in section 5.

	\subsection{Multinacci sequences and matrices}
	\begin{definition}
		The second order Fibonacci sequence $\{f_{k}\}_{k \geq 0}$ is given by
		\begin{eqnarray}
			f_{k+2} = f_{k} + f_{k+1}, \hspace{.5cm}k\geq 2,~where~~f_{0}=0,~~f_{1}=1. \nonumber
		\end{eqnarray}
	\end{definition}
	\begin{definition}[Multinacci Sequence]\cite{prasad2021cryptography}
		For $n(n\geq 2) \in \mathbb{N}$, the generalized Fibonacci sequence\{$t_{k}$\} of order $n$ is given by  
		\begin{equation}\label{genFeb}
			t_{k+n}=t_{k}+t_{k+1}+t_{k+2}+...+t_{k+n-1},~~k\geq 0
		\end{equation}
		where $t_{0}=t_{1}=...= t_{n-2}=0$ and $t_{n-1}=1$. The generalized Fibonacci sequence \{$t_{k}$\} is called  the multinacci sequence.
	\end{definition}
	 Throughout the paper, we use the notation \{$t_{n,k}$\}  to represent the kth term of the Multinacci sequence (generalized Fibonacci sequence) of order n.

	The k-th generalized Fibonacci matrix $F_{n}^k$ (also known as Multinacci matrix) of order n associated with the sequence \{$t_{n,k}$\} is given by
	\begin{equation}\label{kthpower}
		F_{n}^{k}=
		\begin{bmatrix}
			t_{n,k+n-1} & t_{n,k+n-2}+t_{n,k+n-3}+...+t_{n,k} & \cdots &  t_{n,k+n-2} &\\
			t_{n,k+n-2} & t_{n,k+n-3}+t_{n,k+n-4}+...+t_{n,k-1} & \cdots & t_{n,k+n-3} \\ 
			\vdots & \vdots & \ddots & \vdots\\
			t_{n,k+1} & t_{n,k}+t_{n,k-1}+...+t_{n,k-n+2} & \cdots & t_{n,k} \\
			t_{n,k} & t_{n,k-1}+t_{n,k-2}+...+t_{n,k-n+1} & \cdots & t_{n,k-1} \\ 
		\end{bmatrix}_{n \times n},
		\text{where}~k=0, \pm1, \pm2, ...
	\end{equation}
	and
	\begin{eqnarray}
		F_{n}=F_{n}^1=
		\begin{bmatrix}
			t_{n,n} & t_{n,n-1}+t_{n,n-2}+...+t_{n,1} & \cdots &  t_{n,n-1} &\\
			t_{n,n-1} & t_{n,n-2}+t_{n,n-3}+...+t_{n,0} & \cdots & t_{n,n-2} \\ 
			\vdots & \vdots & \ddots & \vdots\\
			t_{n,2} & t_{n,1}+t_{n,0}+...+t_{n,-n+3} & \cdots & t_{n,1} \\ 
			t_{n,1} & t_{n,0}+t_{n,-1}+...+t_{n,-n+2} & \cdots & t_{n,0} \\ 
		\end{bmatrix}
		_{n \times n}=
		\begin{bmatrix}	
			1 & 1 & 1 & ...& 1 & 1\\	
			1 & 0 & 0 & ...& 0 & 0 \\
			\vdots & \vdots & \vdots & \ddots & \vdots & \vdots  \\
			0 & 0 & 0 & ...& 0 & 0 \\
			0 & 0 & 0 &...& 1 & 0 \\
		\end{bmatrix}.\nonumber
	\end{eqnarray}
	
	For $ n=2,3 $ in eqn.(\ref{genFeb}), we get the Fibonacci and Tribonacci sequences respectively and the associated matrix is known as Fibonacci and Tribonacci matrices respectively.
	
	The following lemma lists some properties of the Multinacci matrices $F_{n}^k$, which we use in this paper to establish some results.	
	\begin{lemma}\cite{prasad2021cryptography}\label{lemma11}
		Let $n(\geq 2)\in\mathbb{N}$ and $k,l\in\mathbb{Z}$, then for Multinacci matrices $F_{n}^k$, we have	
		\begin{enumerate}
			\item $F_{n}^{0}=I_{n}$, where is $I_{n}$ the identity matrix of order n
			\item ${(F_n^{1})^k}=F_n^{k}$ and ${(F_n^{-1})^k}=F_n^{-k}$
			\item ${F_n^{k}F_n^{l}}=F_n^{k+l}$ and $F_n^{k}F_n^{-k} = I_{n}$
			\item $det(F_n^{k}) = (-1)^{(n-1)k}$
		\end{enumerate}
	\end{lemma}
	\paragraph{Inverse of Multinacci matrices:} From (4) of lemma(\ref{lemma11}), it has been noted that the determinant of $F_{n}^{k}$ never vanishes for any n and k, hence $F_{n}^{k}$ is always non-singular. Thus, the inverse of the multinacci matrix $F_{n}^{k}$ is given by $F_{n}^{-k}$ which can be achieved by  replacing $k$ by $-k$ in eqn.(\ref{kthpower}).
	\paragraph{Commutative nature:} From (3) of lemma(\ref{lemma11}), it is clear that Multinacci matrices are commutative with respect to usual multiplication.
\section{Multinacci block matrices }
	\begin{definition}[Block Matrix]
		Let $F_n^{m_1}, F_n^{m_2}$ be any two multinacci matrix of order $n$ and $C$ be any square matrix of same order, then the multinacci block matrix(say $A$, in short MBM) is defined as
		\begin{equation}
		A= \begin{bmatrix}
			F_n^{m_1} & C\\
			0 &  F_n^{m_2}\\
			\end{bmatrix}_{2n \times 2n}\nonumber
		\end{equation}
	\end{definition}
	The following lemma deals with the powers of A involving multinacci matrices, which follows a certain pattern.	 
	\begin{lemma}
		For $j \in \mathbb{N} \cup \{0\} $, we have 
		\begin{eqnarray}\label{blockmatrix}
			 A^j=
			\begin{bmatrix}
				F_n^{jm_1} & C^{(j)}\\
				0 &  F_n^{jm_2}\\
			\end{bmatrix}, 	~~where  ~~ C^{(j)}=\begin{cases}
			O_n, & j=0\\
			\sum_{r=0}^{j-1}(F_n^{m_1})^{j-1-r}C(F_n^{m_2})^{r}, & j\geq 1
		\end{cases}
		\end{eqnarray}
	\end{lemma}
	\begin{proof}
	We prove it using mathematical induction on $j$. For $j=1$, we have 
	$ A^1=
	\begin{bmatrix}
		F_n^{m_1} & C^{(1)}\\
		0 &  F_n^{m_2}\\
	\end{bmatrix} =A $ and $C^{(1)} = C$, satisfied.
	\\Next, assuming the statement is true for $j$ (i.e. eqn.(\ref{blockmatrix}) holds), we prove it for $j+1$. Here we have
	\begin{eqnarray}
		A^{j+1}= A^{j}A^{1} &=&
	\begin{bmatrix}
		F_n^{jm_1} & C^{(j)}\\
		0 &  F_n^{jm_2}\\
	\end{bmatrix} 
	\begin{bmatrix}
	F_n^{m_1} & C\\
	0 &  F_n^{m_2}\\
	\end{bmatrix} = 
	\begin{bmatrix}
	F_n^{jm_1}F_n^{m_1} & F_n^{jm_1}C+C^{(j)}F_n^{m_2}\\
	0 &  F_n^{jm_2}F_n^{m_2}\\
	\end{bmatrix}. \nonumber
	\end{eqnarray}
Since, \begin{eqnarray}
	F_n^{jm_1}C+C^{(j)}F_n^{m_2} &=& F_n^{jm_1}C + \left[\sum_{r=0}^{j-1}(F_n^{m_1})^{j-1-r}C(F_n^{m_2})^{r}\right]F_n^{m_2}\nonumber\\
	&=& F_n^{jm_1}C + \left[(F_n^{m_1})^{j-1}C(F_n^{m_2})+ (F_n^{m_1})^{j-2}C(F_n^{m_2})^2 + ...+ C(F_n^{m_2})^j\right]\nonumber\\
	&=& \sum_{r=0}^{j}(F_n^{m_1})^{j-r}C(F_n^{m_2})^{r}  = C^{(j+1)} \nonumber
	\end{eqnarray}
Hence, 
	\begin{eqnarray}
		A^{j+1} = 
	\begin{bmatrix}
		F_n^{(j+1)m_1} & C^{(j+1)}\\
		0 &  F_n^{(j+1)m_2}\\
	\end{bmatrix}, ~
	where~~C^{(j+1)}=\sum_{r=0}^{j}(F_n^{m_1})^{j-r}C(F_n^{m_2})^{r}.\nonumber
	\end{eqnarray}\\
	This completes the proof.
	\end{proof}
	In order to use a matrix as key element in cryptography, it should be necessarily invertible. In our case, we use a encryption method analogs to extended Hill cipher under the prime residue and a part of Multinacci block matrix(MBM, in short) as  key element . To show that MBMs are non singular, it is sufficient to prove that determinant of MBMs are non zero, which has proven in the following lemma.
	\begin{lemma}[Determinant of $ A$]
		The determinant of Multinacci block matrices are
		$$ Det(A)= (-1)^{(n-1)(m_1+m_2)},~\text{where n is order of Multinacci matrices}~F_n^{m_1}~\&~F_n^{m_2}.$$
	\end{lemma}
	\begin{proof} 
		To prove the statement, we use the fact that determinant of a block matrix  $ \begin{bmatrix}
			X & Y\\
			0 &  Z\\
		\end{bmatrix}$ is given by
		$$ 
		det\left(\begin{bmatrix}
			X & Y\\
			0 &  Z\\
		\end{bmatrix}\right)=Det(X)Det(Y). $$
		Therefore, 
		\begin{eqnarray}
			\left(\begin{bmatrix}
				F_n^{m_1} & C\\
				0 &  F_n^{m_2} \\
			\end{bmatrix}\right)
		&=& Det(F_n^{m_1})Det(F_n^{m_2}) \nonumber\\
		&=& (-1)^{(n-1)m_1}(-1)^{(n-1)m_2}\nonumber\\
		&=&(-1)^{(n-1)(m_1+m_2)}.\nonumber
		\end{eqnarray} This complete the proof.
	\end{proof}
	Thus, MBMs are non singular and therefore inverse of MBMs exist. The following lemma gives the inverse of MBMs.
  \begin{lemma}
  	For any Multinacci matrix $F_n^{k}$, the inverse of Multinacci block matrix is given by 
  $$ \begin{bmatrix}
  	F_n^{m_1} & C\\
	  	0 &  F_n^{m_2} \\
	  \end{bmatrix}^{-1} =  
	\begin{bmatrix}
	  F_n^{-m_1} & -F_n^{-m_1}CF_n^{-m_2}\\
	  0 &  F_n^{-m_2} \\
	\end{bmatrix}.$$
	\end{lemma}
	\begin{proof}
		It can be easily proved by using the fact that if B is inverse of A then AB = BA = I.
	\end{proof}	
	Next, for both parties(sender and receiver) to be agree on the same key element(matrix), we have to prove $[C^{(i)}]^{(j)}=[C^{(j)}]^{(i)}$, for which we use the following lemma.
  \begin{lemma}
  	For $i, j\in\mathbb{N}$, we have
  	\begin{eqnarray}
  		\begin{bmatrix}
  			F_n^{m_3} & C^{(j)}\\
  			0 & F_n^{m_4}\\
  		\end{bmatrix}^i
  		=\begin{bmatrix}
  			F_n^{im_3} & [C^{(j)}]^{(i)}\\
  			0 & F_n^{im_4}\\
  		\end{bmatrix}, 	\text{where}~~ [C^{(j)}]^{(i)}=\sum_{s=0}^{i-1}(F_n^{m_3})^{i-1-s}C^{(j)}(F_n^{m_4})^{s},\nonumber \\\begin{bmatrix}
  			F_n^{m_1} & C^{(i)}\\
  			0 & F_n^{m_2}\\
  		\end{bmatrix}^j=
  		\begin{bmatrix}
  			F_n^{jm_1} & [C^{(i)}]^{j}\\
  			0 & F_n^{jm_2}\\
  		\end{bmatrix}, \text{where}~~ [C^{(i)}]^{(j)}=\sum_{r=0}^{j-1}(F_n^{m_1})^{j-1-r}C^{(i)}(F_n^{m_2})^{r}.\nonumber
  	\end{eqnarray}
  \end{lemma}
  \begin{theorem}\label{matrtixcom}
  	$[C^{(i)}]^{(j)}=[C^{(j)}]^{(i)}$ for all $ i,j\in\mathbb{N} $.
  \end{theorem}
  \begin{proof}
  	By using the property of commutativity of the multinacci matrices, we have
  	\begin{eqnarray}
  		[C^{(i)}]^{(j)}&=&\sum_{r=0}^{j-1}(F_n^{m_1})^{j-1-r}C^{(i)}(F_n^{m_2})^{r}\nonumber\\
  		&=&\sum_{r=0}^{j-1}(F_n^{m_1})^{j-1-r}\left[\sum_{s=0}^{i-1}(F_n^{m_3})^{i-1-s}C(F_n^{m_4})^{s}\right](F_n^{m_2})^{r}\nonumber\\
  		&=&\sum_{r=0}^{j-1}\sum_{s=0}^{i-1}(F_n^{m_1})^{j-1-r}(F_n^{m_3})^{i-1-s}C(F_n^{m_4})^{s}(F_n^{m_2})^{r}\nonumber\\
  		&=&\sum_{r=0}^{j-1}\sum_{s=0}^{i-1}(F_n^{m_3})^{i-1-s}(F_n^{m_1})^{j-1-r}C(F_n^{m_2})^{r}(F_n^{m_4})^{s}\nonumber\\
  		&=&\sum_{s=0}^{i-1}(F_n^{m_3})^{i-1-s}\left[\sum_{r=0}^{j-1}(F_n^{m_1})^{j-1-r}C(F_n^{m_2})^{r}\right](F_n^{m_4})^{s}\nonumber\\
  		&=&\sum_{s=0}^{i-1}(F_n^{m_3})^{i-1-s}C^{(j)}(F_n^{m_4})^{s}\nonumber\\
  		&=&[C^{(j)}]^{(i)}.\nonumber
  	\end{eqnarray}
  As required.
  \end{proof}
	\subsection{Key Generation Algorithm}
	In cryptography, elements of key component plays a crucial role for efficient encryption and better security. We are using Fibonacci block matrices for key composition \& encryption and vice-versa. It has been discussed below in the steps followed by encryption algorithm.\\
	Let us consider $S=\left\{ \begin{bmatrix}
		F_n^{m_1}(\mathbb{Z}_p) & K\\
		0 &  F_n^{m_2}(\mathbb{Z}_p)\\
	\end{bmatrix} :K \in M_{n}(\mathbb{F}_{p})\right\}$.
	where we used the notation $ F_n^{k}(\mathbb{Z}_p) $ for Multinacci matrices over $\mathbb{Z}_p$ and $M_{n}(\mathbb{F}_{p})$ for matrices over $\mathbb{Z}_p$.
\subsection{Construction of public key}
	Let us assume that the communication is being made between two parties, Alice and Bob. So, for public key setup, Alice do the following steps.
	\begin{enumerate}
	\item Alice chooses a prime number $p$, $ l\in \mathbb{N} $ and matrix $A= \begin{bmatrix}
		G & K\\
		0 &  H\\
	\end{bmatrix} \in S $ with $G= F_n^{m_1}(\mathbb{Z}_p)$ and $H=F_n^{m_2}(\mathbb{Z}_p)$.

		\item Calculate key element $ K^{(l)} $ as
		$$K^{(l)}= \sum_{r=0}^{l-1}(G)^{l-1-r}K(H)^{r}\pmod p.$$\\
		Now, Alice makes $(p ,K,K^{(l)})$ as public key and keep $(G,H, l)$ as her secret key.
	\end{enumerate}
\subsection{Key Generation and Encryption}
	Let P represent the plaintext partitioned as $P= (P_1P_2...P_n).$ and C is the corresponding ciphertext $C= (C_1C_2...C_n)$. Now, using Alice's public key $(p ,K,K^{(l)})$, Bob generates his  encryption key and then encrypts his plaintext as follows:
	\begin{enumerate}
		\item Bob chooses a secret key, say $j \in \mathbb{N} $ and $B= \begin{bmatrix}
			M & K\\
			0 & N\\
		\end{bmatrix} \in S $ with $M= F_n^{m_3}(\mathbb{Z}_p)$ and $N=F_n^{m_4}(\mathbb{Z}_p).$
		\item Calculate, $K^{(j)}=\sum_{s=0}^{j-1}(M)^{j-1-s}K(N)^{s} \pmod p.$
		\item Calculate encryption key as 
		\begin{equation}\label{keymatrix}
			 [K^{(l)}]^{(j)}=\sum_{s=0}^{j-1}(M)^{j-1-s}K^{(l)}(N)^{s} \pmod p.
		\end{equation}
	  Thus, his encryption key (say $E_{k})$ is $ [K^{(l)}]^{(j)}$.
		\item Now, Bob construct a row vector E of size n over field $\mathbb{F}_{p}$ whose $ i^{th} $ column is the sum of elements of $ i^{th} $ column of $E_{K}$ .
		\item Encryption method: $ C_i \equiv (P_iE_{K}+E) \pmod p $, where $C= (C_1C_2...C_n).$
		\item Finally, Bob sends $(K^{(j)},C)$ to Alice.\\
	\end{enumerate}
\subsection{Decryption}
	On the other side, after receiving $(K^{(j)},C)$ from Bob, Alice perform following operations to recover the plaintext:
	\begin{enumerate}
		\item Alice first calculate key matrix $E_{k}$ as, $$E_{k}=[K^{(j)}]^{(l)}=\sum_{r=0}^{l-1}(G)^{l-1-r}*K^{(j)}*(H)^{r} \pmod p.$$
		\item Thus, decryption key (say D$_{K}$) = $(E_{k})^{-1}$.
		\item Decryption of ciphertext: $ P_i \equiv (C_i-E)D_{K} \pmod p $
		where E is a row vector over $\mathbb{F}_{p}$ whose $ i^{th} $ column is the sum of elements of $ i^{th} $ column of $[K^{(j)}]^{(l)}$ over $\mathbb{F}_{p}$.
	\end{enumerate}
	The above methodology is illustrated by an example in the following section.
	\section{Numerical example}
	\begin{example}
		Consider $p=47$ and $S=\left\{ \begin{bmatrix}
			F_3^{m_1}(\mathbb{Z}_p) & K\\
			0 &  F_3^{m_2}(\mathbb{Z}_p)\\
		\end{bmatrix} :K \in M_{3}(\mathbb{F}_{p})\right\}$. Encrypt the plaintext \textbf{HEY} using proposed method.
	\end{example}
	\begin{proof}[Solution]
		Assume, Bob wish to send a plaintext \textbf{HEY} to Alice. So for encryption, Bob need public key of Alice.
		\paragraph{Construction of Alice's Public Key:}
	\begin{enumerate}
		\item Alice chooses a random number, say $l = 5$ and $A=
		\begin{bmatrix}
			G & K\\
			0 &  H\\
		\end{bmatrix} \in S$, where
		\begin{eqnarray}
			G= F_{3}^{9}=
			\begin{bmatrix}
				8 & 31 & 34\\
				34 & 21 & 44\\
				44 & 37 & 24\\
			\end{bmatrix},
			H= F_{3}^{13}
			\begin{bmatrix}
				13 & 21 & 34\\
				34 & 26 & 34\\
				34 & 0 & 9\\
			\end{bmatrix}
			\text{and}~ 
			K= 
			\begin{bmatrix}
				2 & 3 & 1\\
				1 & 1 & 1\\
				1 & 0 & 0\\
			\end{bmatrix}.\nonumber
		\end{eqnarray} 
		\item Construction of public key,
		$$K^{(5)}=\sum_{r=0}^{4}(G)^{4-r}K(H)^{r} \pmod {47}
		\equiv 
		\begin{bmatrix}
			42 & 25 & 5\\
			5 & 37 & 20\\
			20 & 32 & 17\\
		\end{bmatrix}.$$
	\end{enumerate}
	Thus, Alice's public key is
	$\left( 47,
	\begin{bmatrix}
		2 & 3 & 1\\
		1 & 1 & 1\\
		1 & 0 & 0\\
	\end{bmatrix},
	\begin{bmatrix}
		42 & 25 & 5\\
		5 & 37 & 20\\
		20 & 32 & 17\\
	\end{bmatrix}\right)$ and her secret key is\\
	$\left( 5,
	\begin{bmatrix}
		8 & 31 & 34\\
		34 & 21 & 44\\
		44 & 37 & 24\\
	\end{bmatrix},
	\begin{bmatrix}
		13 & 21 & 34\\
		34 & 26 & 34\\
		34 & 0 & 9\\
	\end{bmatrix}\right)$.\\\\
	\paragraph{Key Generation and Encryption(Bob side):}
	Now, Bob construct his encryption key using Alice’s\vspace{.3cm} public key   $\left( 47,
	\begin{bmatrix}
		2 & 3 & 1\\
		1 & 1 & 1\\
		1 & 0 & 0\\
	\end{bmatrix},
	\begin{bmatrix}
		42 & 25 & 5\\
		5 & 37 & 20\\
		20 & 32 & 17\\
	\end{bmatrix}\right)$\vspace{.3cm},  and encrypt his plaintext as follows:\\
	\begin{enumerate}
		\item Row vector of plaintext: $P \leftarrow [07,04, 24]$~~~ $(Here, [H,E,Y]=[07,04,24]).$
		\item Bob chooses his secret number, say $m=3$ and matrix $B \in S$ with
		\begin{eqnarray}
			M = F_{3}^{7} = 
			\begin{bmatrix}
				44 & 37 & 24\\
				24 & 20 & 13\\
				13 & 11 & 7\\
			\end{bmatrix}
			\text{and}~~
			N= F_{3}^{15}=
			\begin{bmatrix}
				34 & 0 & 34\\
				34 & 0 & 13\\
				13 & 21 & 34\\
			\end{bmatrix}.\nonumber
		\end{eqnarray}
		\item Calculate $$K^{(3)}=\sum_{s=0}^{2}(M)^{2-s}*K*(N)^{s}  \pmod {47} 
		= 
		\begin{bmatrix}
			24 & 4 & 19\\
			19 & 5 & 32\\
			32 & 34 & 20\\
		\end{bmatrix}.$$
		\item Calculation of encryption key, $$[K^{(5)}]^{3}=\sum_{s=0}^{2}(M)^{2-s}K^{(5)}(N)^{s} \pmod {47}=
		\begin{bmatrix}
			34 & 19 & 5\\
			5 & 29 & 14\\
			14 & 38 & 15\\
		\end{bmatrix}.$$\\ 
		Thus, encryption key $E_{K} =  [K^{(5)}]^{3}$ .
		
		\item Bob's row vector E over $\mathbb{Z}_{47}$ is $E= 
		\begin{bmatrix}
			6 & 39 & 34\\
		\end{bmatrix} $.
		
		\item Encryption:  $C\equiv (PE_{K}+E) \pmod {47}$ .\vspace{.3cm}\\
		$C \equiv \left( 
		\begin{bmatrix}
			7 & 4 & 24\\
		\end{bmatrix}*
		\begin{bmatrix}
			34 & 19 & 5\\
			5 & 29 & 14\\
			14 & 38 & 15\\
		\end{bmatrix}+
		\begin{bmatrix}
			6 & 39 & 34\\
		\end{bmatrix}\right) \pmod {47}$  \vspace{.3cm}
		\\$~~~~\equiv\begin{bmatrix}
			36 & 25 & 15\\
		\end{bmatrix}$\vspace{.2cm} $\rightarrow$ [$>$~~Z~~P].\vspace{.3cm}\\
		Here, plaintext $[HEY]$ encrypted as $[>ZP]$.
		
		\item Bob sends $\left( C, K^{(3)}\right)$ = 
		$\left([>ZP], 	
		\begin{bmatrix}
			24 & 4 & 19\\
			19 & 5 & 32\\
			32 & 34 & 20\\
		\end{bmatrix}\right)$ to Alice.	
	\end{enumerate}
	\paragraph{Decryption (Alice side):}
	After receiving $\left( C, K^{(3)}\right) = 
	\left(>ZP, 	
	\begin{bmatrix}
		24 & 4 & 19\\
		19 & 5 & 32\\
		32 & 34 & 20\\
	\end{bmatrix}\right)$ from Bob,\\ Alice performs the following steps to recover the plaintext.
	\begin{enumerate}
		\item Alice calculates key matrix as   
		$$E_{K}=[K^{(3)}]^{(5)}=\sum_{r=0}^{4}(G)^{4-r}*K_{3}*(H)^{r} =
		\begin{bmatrix}
			34 & 19 & 5\\
			5 & 29 & 14\\
			14 & 38 & 15\\
		\end{bmatrix}.$$
		\item Decryption key D$_{K}$ over $\mathbb{Z}_{47}$ is $(E_{K})^{-1}= 
		\begin{bmatrix}
			43 & 30 & 36\\
			36 & 7 & 41\\
			41 & 42 & 13\\
		\end{bmatrix}$.
		
		\item Decryption method: $P \equiv (C-E)D_{K} \pmod {47}$.
		\\Here, $C=[>,Z,P] \rightarrow$ [36~~25~~15] and row vector 
		$E= \begin{bmatrix}
			6 & 39 & 34\\
		\end{bmatrix}$.
		Thus, Alice recovers the plaintext as
		\begin{eqnarray}
			P 	&\equiv& \left(\left(\begin{bmatrix}
				36 & 25 & 15\\
			\end{bmatrix}-
			\begin{bmatrix}
				6 & 39 & 34\\
			\end{bmatrix}\right)*
			\begin{bmatrix}
				43 & 30 & 36\\
				36 & 7 & 41\\
				41 & 42 & 13\\
			\end{bmatrix}\right) \pmod {47}\nonumber\\
			&\equiv&
			\begin{bmatrix}
				7 & 4 & 24\\
			\end{bmatrix}\to [H~~E~~Y].\nonumber
		\end{eqnarray}
	\end{enumerate}	
	Thus, the message \textbf{HEY } successfully reached to Alice.
	\end{proof} 
\section{Key space and mathematical strength}	
	Security strength of proposed scheme depends on the computational power require to achieve the private key $(j,M,N)$ of sender and private key $(l,G,H)$ of receiver.
	Our, encryption key is formulated as
	\begin{eqnarray}
		[K^{(l)}]^{(j)}=\sum_{s=0}^{j-1}(M)^{j-1-s}*K^{(l)}*(N)^{s} \nonumber
	\end{eqnarray}
	and after encryption Bob transmits $(K^{(j)}, C)$ to Alice through a unsecure channel. So, we assume that intruder may know $(K^{(j)}, C)$ by unfair means but after knowing $(K^{(j)}, C)$, intruder needs matrices M, N to calculate encryption key $[K^{(l)}]^{(j)}$. Since there is no any deterministic polynomial time algorithm (Discrete logarithm problem \cite{hoffstein2008introduction,stinson2005cryptography}) to calculate $M, N$ from $[K^{(l)}]^{(j)}$ so, it is almost impossible to recover encryption key from given information on large primes.

	Key space based on assumed parameters follows from matrix theory.
	In matrix theory $GL_{n}(\mathbb{F}_{p})$ represents the set of invertible matrices of order $n\times n$ over finite field $\mathbb{F}_{p}$, where p is an odd prime. The order of General Linear group ($GL_{n}$) over finite field $\mathbb{F}_{p}$ is given by
	\begin{equation}\label{GLN}
		|GL_{n}(\mathbb{F}_{p})| = (p^{n}-p^{n-1})(p^{n}-p^{n-2}) \cdots (p^{n}-p^{1})(p^{n}-1).
	\end{equation}
	To examine strength of our key space, we are presenting a table of possible key spaces over $\mathbb{F}_{p}$ based on General Linear group. For simplicity, considering matrices of order 3 × 3 and 4 × 4.
	\begin{center}
		\begin{tabular}{|c|c|c|c|}
			\hline 
			Prime(p) & Possible Key spaces on $GL_{3}(\mathbb{F}_{p})$ & Possible Key spaces on $GL_{4}(\mathbb{F}_{p})$ \\
			\hline \hline
			3 & 1.1232$\times 10^{4}$ & 2.4261$\times 10^{7}$\\
			\hline
			5 & 1.4880$\times 10^{6}$ & 1.1606$\times 10^{11}$\\ 
			\hline
			7 & 3.3784$\times 10^{14}$ & 2.7811$\times 10^{13}$\\
			\hline
			11 & 3.1920$\times 10^{9}$ & 6.2166$\times 10^{25}$\\ 
			\hline
			13 & 9.7264$\times 10^{9}$ & 6.1029$\times 10^{18}$\\
			\hline
			17 & 1.0948$\times 10^{11}$ & 4.5630$\times 10^{19}$\\
			\hline
			19 & 3.0481$\times 10^{11}$ & 2.7246$\times 10^{20}$\\ 
			\hline
			23 & 1.7194$\times 10^{12}$ & 5.8543$\times 10^{21}$\\
			\hline
			29 & 1.1499$\times 10^{16}$ & 3.6139$\times 10^{28}$\\
			\hline	
			\vdots & \vdots & \vdots \\ 
			\hline
		\end{tabular}
	\end{center}
	From the table, it has been observed that the growth rate of key space is very high when $p \rightarrow \infty $. Thus, we conclude that if size of key matrix is increasing along with prime, then it forms a very large key space which can be easily done with Fibonacci matrices.
\section{Conclusion}
     We have proposed multinacci block matrices, a type of upper triangular matrix involving multinacci matrices at diagonal places. Moreover, we have obtained some algebraic properties of multinacci block matrices and proposed a public key cryptography using it.
     Our proposed cryptography is based on the key element from block matrices over a ﬁnite ﬁeld $\mathbb{F}_{p}$.
     Here, we have used the multiplicative commutativity of the multinacci matrices for agreement of end users on the same key.
     
     Here, the set $S=\left\{ \begin{bmatrix}
     	F_n^{m_1}(\mathbb{Z}_p) & K\\
     	0 &  F_n^{m_2}(\mathbb{Z}_p)\\
     \end{bmatrix} :K \in M_{n}(\mathbb{F}_{p})\right\}$ is global element i.e. known to everyone, a sender can choose any matrix from the set $S$ to construct the key matrix. So in the above proposed scheme, neither sender nor receiver needs to publish their corresponding set of matrices that commute with a chosen matrix of sender and receiver respectively. Our proposed schemes has a large key space and its security relies on the discrete logarithm problems.
\section*{Acknowledgment}		
	The first author acknowledge the University Grant Commission(UGC), India for providing fellowship for this research work.

	\bibliography{References}
	\bibliographystyle{acm}
	
\end{document}